\documentclass{svproc}
\usepackage{url}
\usepackage{graphicx}
\usepackage{bm}

\begin{document}
\mainmatter              % start of a contribution
\title{Transport coefficients of hot and dense matter}
%
%\titlerunning{Hamiltonian Mechanics}  % abbreviated title (for running head)
\author{O. Soloveva\inst{1} \and P. Moreau\inst{1}
L. Oliva\inst{2,1} \and T. Song\inst{2} \and W. Cassing\inst{3} \and \\E. Bratkovskaya\inst{2,1}}
%
%\authorrunning{Ivar Ekeland et al.} % abbreviated author list (for running head)
\tocauthor{Ivar Ekeland, Roger Temam, Jeffrey Dean, David Grove,
Craig Chambers, Kim B. Bruce, and Elisa Bertino}
\institute{Institute for Theoretical Physics, Johann Wolfgang
Goethe-Universit\"{a}t, Frankfurt am Main, Germany,\\
\email{soloveva@fias.uni-frankfurt.de},\\
\and
GSI
Helmholtzzentrum f\"{u}r Schwerionenforschung GmbH,
 Darmstadt, Germany
 \and
Institut f\"{u}r Theoretische Physik, Universit\"{a}t
Gie\ss en, Germany}

\maketitle              % typeset the title of the contribution

\begin{abstract}
We present calculations for the shear viscosity of the hot and dense quark-gluon plasma (QGP) using the partonic scattering cross sections as a function of temperature $T$ and baryon chemical potential $\mu_B$ from the dynamical quasiparticle model (DQPM) that is matched to reproduce the equation of state of the partonic system above the deconfinement temperature $T_c$ from lattice QCD. To this aim we calculate the collisional widths for the partonic degrees of freedom at finite $T$ and $\mu_B$ in the time-like sector and conclude that the quasiparticle limit holds sufficiently well. Furthermore, the ratio of shear viscosity $\eta$ over entropy density $s$, i.e.  $\eta/s$, is evaluated using these collisional widths and are compared to lQCD calculations for $\mu_B$ = 0 as well. We find that the ratio $\eta/s$ is in agreement with the results of calculations within the original DQPM on the basis of the Kubo formalism. Furthermore, there is only a very modest change of $\eta/s$ with the baryon chemical $\mu_B$ as a function of the scaled temperature $T/T_c(\mu_B)$.
\end{abstract}

\section{Introduction}

Transport coefficients of the hot and dense QGP are important ingridients for a fundamental description of medium properties. An exploration of temperature $T$ and baryon chemical potential $\mu_B$ dependences of transport coefficients will provide useful information for the hydrodynamical simulations of heavy-ion collisions (HICs). It has been found that the QGP, produced in the central regions of HICs at the Relativistic Heavy Ion Collider (RHIC), is a strongly interacting system. The experimantal data for elliptic flow can be well reproduced by hydrodynamical simulations with a small value of the shear viscosity over entropy density \cite{Romatschke,Song:Heinz}. 
Calculations in the vicinity to the deconfinement phase transition $T_c$ on the basis of perturbative QCD are notoriously difficult since the coupling rises in this region. In the strong interacting regime the methods of lattice gauge theories can be applied. Results from the lattice QCD calculations available only for $\mu_B$=0, so far the presence of a non-zero chemical potential $\mu_B$ might affect the values of transport coefficients substantially. In light of these difficulties the evaluations of transport coefficients for a non-zero $\mu_B$ are necessary. One can evaluate transport coefficients for a non-zero $\mu_B$ within the effective approaches, which are found to match well the lQCD equation of state. We will present calculations for the shear viscosity within the DQPM for moderate values of $\mu_B \leq $ 450 MeV, where we assume that in this region the transition between the hadronic and the quark-gluon plasma phase is a smooth cross-over.

\section{Dynamical quasiparticle model (DQPM)}
We describe the QGP in equilibrium on the base of the dynamical quasiparticle model (DQPM), which is based on partonic propagators with sizable
imaginary parts of the selfenergies incorporated \cite{Review}. Whereas the real part of the self-energies can be attributed to a dynamically generated mass (squared) the imaginary parts contain the information about the interaction rates in the system. Furthermore,
the imaginary parts of the propagators define the spectral functions of the degrees of freedom
which might show quasiparticle peaks. A further advantage of a propagator-based approach is that one can formulate a consistent thermodynamics \cite{Baym} as well as a causal theory for non-equilibrium configurations on the basis of the Kadanoff-Baym equations.
 In order to explore the transport properties of a partonic system we have calculated the interaction rates using the partonic differential cross sections as a function of $T$ and $\mu_B$, which are evaluated for the leading tree-level diagrams using the DQPM (cf. Appendices of Ref. \cite{Pierre19}). In the on-shell case (energies of the particles are taken to be $E^2 = \mathbf{p}^2 + M^2$ with $M$ being the pole mass) the interaction rate $\Gamma_i^{on}$ is obtained as follows:
\begin{equation}
\Gamma_i^{on}  (p_i, T,\mu_B) = \frac{1}{2E_i} \sum_{j=q,\bar{q},g} \int \frac{d^3p_j}{(2\pi)^3 2E_j}\ d_j\ f_j(E_j,T,\mu_B) \int \frac{d^3p_3}{(2\pi)^3 2E_3}
\label{Gamma_on}
\end{equation}
$$ \times  \int \frac{d^3p_4}{(2\pi)^3 2E_4} (1\pm f_3) (1\pm f_4)  |\bar{\mathcal{M}}|^2 (p_i,p_j,p_3,p_4)\ (2\pi)^4 \delta^{(4)}\left(p_i + p_j -p_3 -p_4 \right) \\$$
$$=  \sum_{j=q,\bar{q},g} \int \frac{d^3p_j}{(2\pi)^3}\ d_j\ f_j\ v_{rel} \int d\sigma^{on}_{ij \rightarrow 34}\ (1\pm f_3) (1\pm f_4) , $$
where $d_j$ is the degeneracy factor for spin and color (for quarks $d_q = 2 \times N_c$ and for gluons $d_g =2 \times (N_c^2-1)$), and with the shorthand notation $f_j = f_j(E_j,T,\mu_B)$ for the Fermi and Bose distribution functions. In Eq. (\ref{Gamma_on}) and in all the following sections, the notation $\sum_{j=q,\bar{q},g}$ includes the contribution from all possible partons which in our case are the gluons and the (anti-)quarks of three different flavors ($u,d,s$).

\section{Shear viscosity of the hot and dense QGP}

The starting point to evaluate viscosity coefficients for the partonic matter is the Kubo formalism \cite{Kubo:1957mj} which was also used to calculate the viscosities  within the PHSD in a box with periodic boundary conditions (cf. Ref. \cite{Ozvenchuk:2012kh}). We focus here on the calculation of the shear viscosity $\eta$ based on Ref. \cite{Aarts:2002cc} which reads:
\begin{equation} \label{eta_Kubo}
\eta^{\rm{Kubo}}(T,\mu_B)  = - \int \frac{d^4p}{(2\pi)^4}\ p_x^2 p_y^2
 \sum_{i=q,\bar{q},g} d_i\ \frac{\partial f_i(\omega)}{\partial \omega}\ \rho_i(\omega,\mathbf{p})^2 ,
 \end{equation}
where the notation $f_i(\omega) = f_i(\omega,T,\mu_B)$ is used again for the distribution functions, and $\rho_i$ denotes the spectral function of the partons, while $d_i$ stand for the degeneracy factors. We note that the derivative of the distribution function accounts for the Pauli-blocking (-) and Bose-enhancement (+) factors. Following Ref. \cite{Lang:2012}, we can evaluate the integral over $\omega = p_0$ in Eq. (\ref{eta_Kubo}) by using the residue theorem. When keeping only the leading order contribution in the width $\gamma(T,\mu_B)$ from the residue - evaluated at the poles of the spectral function $\omega_i = \pm \tilde{E}(\mathbf{p}) \pm i \gamma$ - we finally obtain:
\begin{equation} \label{eta_on}
 \eta^{\rm{RTA}}(T,\mu_B)  = \frac{1}{15T} \int \frac{d^3p}{(2\pi)^3}  \sum_{i=q,\bar{q},g}
   \left( \frac{\mathbf{p}^4}{E_i^2 \ \Gamma_i(\mathbf{p}_i,T,\mu_B)}\ d_i \left( (1 \pm f_i(E_i)) f_i(E_i) \right) \right) , \end{equation}
which corresponds to the expression derived in the relaxation-time approximation (RTA) \cite{Sasaki:2008} by identifying the interaction rate $\Gamma$ with $2\gamma$ as expected from transport theory in the quasiparticle limit \cite{Blaizot:1999}. We recall that $\gamma$ is the width parameter in the parton propagator. The interaction rate $\Gamma_i(\mathbf{p}_i,T,\mu_B)$ (inverse relaxation time) here is calculated microscopically using the differential cross sections for parton scattering as described in Section 2.
%We use here the notation $\sum_{j=q,\bar{q},g}$ which includes the contribution from all possible partons which in our case %are the gluons and the (anti-)quarks of three different flavors ($u,d,s$).

\begin{figure}[h]
\begin{minipage}[h]{0.49\linewidth}
\centering
\includegraphics[width=0.97\linewidth]{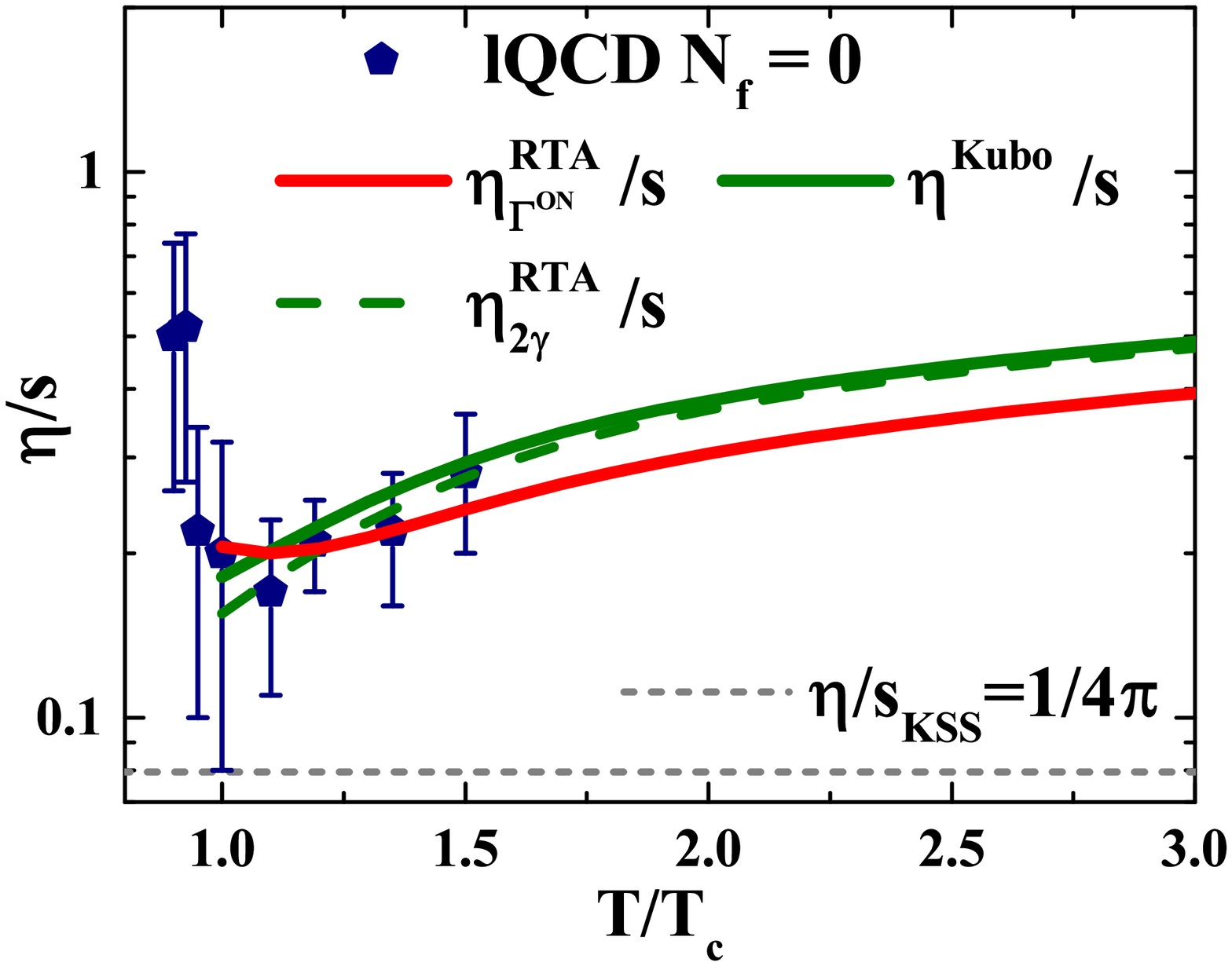}
\end{minipage}
\hfill
\begin{minipage}[h]{0.49\linewidth}
\centering
\includegraphics[width=0.99\linewidth]{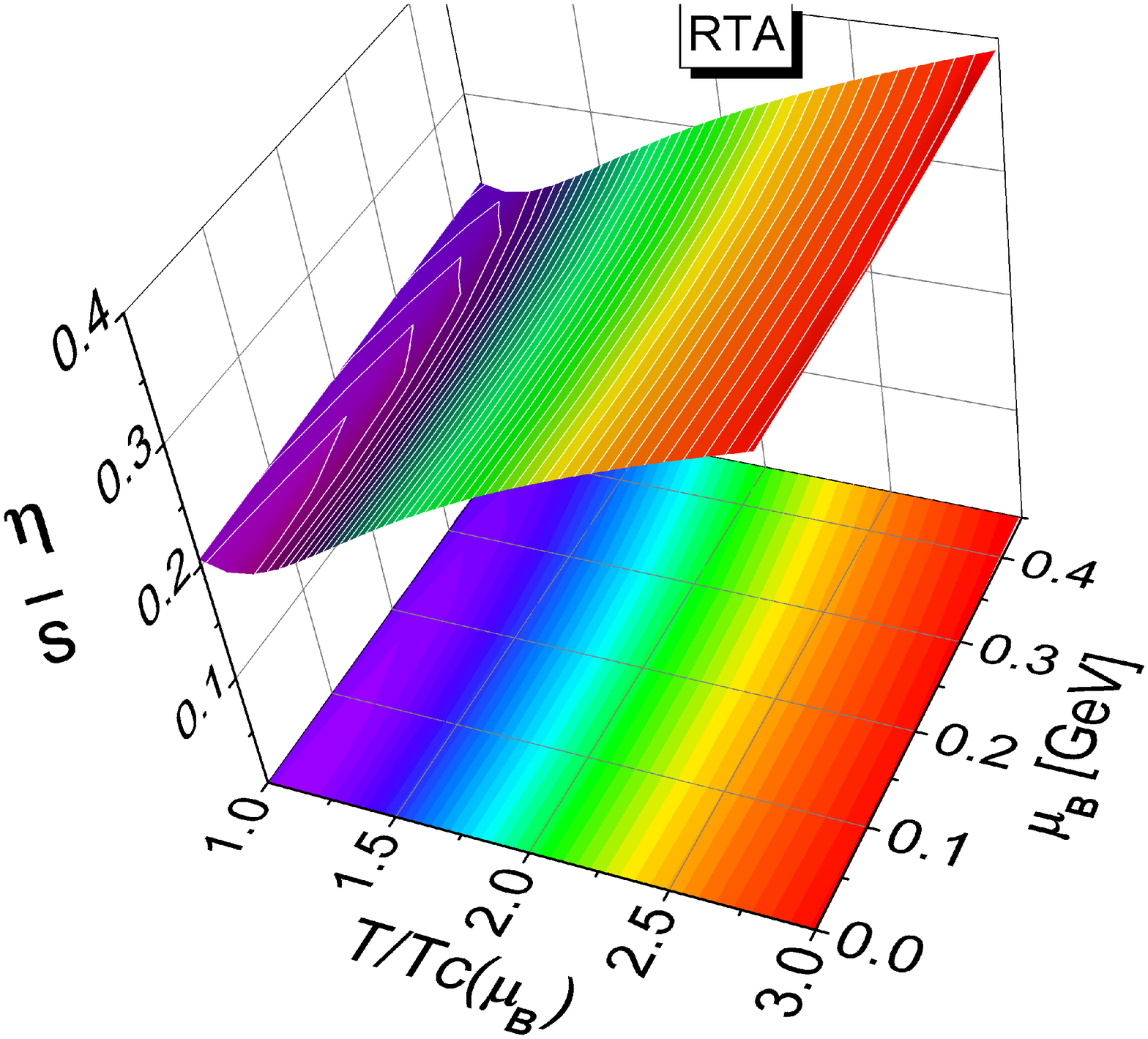}
\end{minipage}
    \caption{(Left) The ratio of shear viscosity to entropy density as a function of the scaled temperature $T/T_c$ for $\mu_B = 0$ from Eq. (\ref{eta_Kubo}-\ref{eta_on}). The solid green line $(\eta^{\rm{Kubo}}/s)$ shows the results from the original DQPM in the Kubo formalism while the dashed green line $(\eta^{\rm{RTA}}_{2\gamma}/s)$ shows the same result in the relaxation-time approximation (\ref{eta_on}). The solid red line $(\eta^{\rm{RTA}}_{\Gamma^{\rm{on}}}/s)$ results from Eq. (\ref{eta_on}) using the interaction rate $\Gamma^{\rm{on}}$  calculated by the microscopic differential cross sections in the on-shell limit. The dashed gray line demonstrates the Kovtun-Son-Starinets bound \cite{Kovtun:2004} $(\eta/s)_{\rm{KSS}} = 1/(4\pi)$. The symbols show lQCD data for pure SU(3) gauge theory taken from Ref. \cite{Astrakhantsev:2017} (pentagons).
(Right) The ratio of shear viscosity to entropy density as a function of the scaled temperature $T/T_c$ from the DQPM for $\mu_B$ in the range  $[0,0.45] $ GeV  from Eq. (\ref{eta_on})}
	\label{fig_eta}
\end{figure}

The actual results are displayed in Fig. \ref{fig_eta} for the
ratio of shear viscosity to entropy density $\eta/s$ as a function
of the scaled temperature $T/T_c$ for $\mu_B$ = 0 in comparison to
those from lattice QCD \cite{Astrakhantsev:2017}.
The solid green line $(\eta^{\rm{Kubo}}/s)$ shows the result from the original DQPM in the
Kubo formalism  while the dashed green line $(\eta^{\rm{RTA}}_{2\gamma}/s)$ shows the same result in
the relaxation-time approximation (\ref{eta_on}) by replacing
$\Gamma_i$ by $2\gamma_i$. The solid red line $(\eta^{\rm{RTA}}_{\Gamma^{\rm{on}}}/s)$ results from Eq.
(\ref{eta_on}) using the interaction rate $\Gamma^{\rm{on}}$
 calculated by the microscopic differential cross
sections in the on-shell limit. We find that the ratios $\eta/s$ do not differ
very much and have a similar behavior as a function of temperature.
The approximation (\ref{eta_on}) of the shear viscosity is found to
be very close to the one from the Kubo formalism (\ref{eta_Kubo})
indicating that the quasiparticle limit ($\gamma \ll M$) holds in the DQPM.

We note in passing that there is no strong variation with $\mu_B$ for fixed $T/T_c(\mu_B)$, however, the ratio increases slightly with $\mu_B$ in the on-shell limit while it slightly drops with $\mu_B$ in the Kubo formalism for the DQPM \cite{Pierre19}. Accordingly, there is some model uncertainty when extracting the shear viscosity in the different approximations.
\subsection*{Acknowledgments}
The authors acknowledge inspiring discussions with J. Aichelin, H. Berrehrah, C. Ratti, E. Seifert, A. Palmese and T. Steinert. Furthermore, P.M., L.O. and E.B. acknowledge support by the DFG grant CRC-TR 211 'Strong-interaction matter under extreme conditions' - Project number 315477589 - TRR 211. O.S. acknowledges support from HGS-HIRe for FAIR; L.O. and E.B. thank the COST Action THOR, CA15213.
The computational resources have been provided by the LOEWE-Center for Scientific Computing.


\end{document}